# *Atmospheric Imaging Assembly* Response Functions: Solving the Fe VIII Problems with *Hinode* EIS Bright Point Data

J.T. Schmelz • B.S. Jenkins • J.A. Kimble

Physics Department, University of Memphis, Memphis, TN 38152, USA
email: jschmelz@memphis.edu

## Abstract

The *Atmospheric Imaging Assembly* (AIA) onboard the *Solar Dynamics Observatory* is a state-of-the-art imager with the potential to do unprecedented time-dependent multi-thermal analysis at every pixel on scales short compared to the radiative and conductive cooling times. Recent results, however, have identified missing spectral lines in the CHIANTI atomic physics data base, which is used to construct the instrument response functions. This is not surprising since the wavelength range from 90 Å to 140 Å has rarely been observed with solar spectrometers, and atomic data for many of these ions are simply not available in the literature. We have done differential emission measure analysis using simultaneous AIA and *Hinode*/EIS observations of six X-ray bright points. Our results not only support the conclusion that CHIANTI is incomplete near 131 Å, but more importantly, suggest that the peak temperature of the Fe VIII emissivity/response is likely to be closer to $\log T = 5.8$ than to the current value of $\log T = 5.7$. Using a revised emissivity/response calculation for Fe VIII, we find that the observed AIA 131-Å flux can be underestimated by $\approx 1.25$, which is smaller than previous comparisons. Making these adjustments brings not only the AIA 131-Å data but also the EIS Fe VIII lines into better agreement with the remainder of the bright point data. In addition, we find that CHIANTI is reasonably complete in the AIA 171- and 193-Å bands. For the AIA 211-, 335-, and 94-Å channels, we recommend that more work be done with AIA–EIS DEM comparisons using observations of active region cores, coronal structures with more warm emission measure than our bright points. Then direct comparisons can be made with a variety of EIS iron lines.

keywords: Sun: corona, Sun: UV radiation, Sun: fundamental parameters

## Introduction

The *Atmospheric Imaging Assembly* (AIA: Lemen *et al.*, 2012) onboard NASA's *Solar Dynamics Observatory* (SDO) is providing images of the solar corona. Each of the EUV channels of AIA is dominated by an iron emission line, each of a different ionization stage, from Fe VIII to Fe XVIII (plus flare lines). This feature not only makes AIA a state-of-the-art imager, but also gives it better temperature resolution than any of its non-spectrometer predecessors. It has the ability to do time-dependent multi-thermal analysis at every pixel on scales short compared to the radiative and conductive cooling times and, therefore, to add significantly to our knowledge of solar coronal physics.

With this goal in mind, several authors have investigated the completeness of the CHIANTI atomic physics data base (www.chiantidatabase.org/referencing.html) in reproducing observed spectra in the AIA coronal bandpasses: Fe VIII (131 Å), Fe IX (171 Å), Fe XII (193 Å), Fe XIV

(211 Å), Fe XVI (335 Å), and Fe XVIII (94 Å). O'Dwyer *et al.* (2010) examined the contributions of spectral lines and continuum emission to each AIA response, and the sensitivity as a function of temperature for each filter. They generated synthetic spectra using CHIANTI and sample differential emission measure (DEM) curves for coronal hole, quiet Sun, active region, and flare plasma. They convolved the synthetic spectra with the effective area of each channel and found the contribution of particular spectral lines, which can affect the interpretation of AIA data under certain conditions.

Schmelz *et al.* (2011a,b) did DEM analysis of non-flaring coronal loop data using AIA. They found several examples where the DEMs had to be extremely broad, with significant emission measure at $\log T \approx 6.8 - 7.0$, in order to give a reasonable DEM fit to the observations. Such high temperatures, however, were inconsistent with the results of spectrometers. They attributed the problems to cool ($\log T \approx 6.0$) spectral lines, which contributed to the observed AIA 131- and 94-Å fluxes, but were not yet part of the CHIANTI database or the instrument response functions (*e.g.* Lepson *et al*, 2002). As a result, too much high-temperature plasma was required in order to obtain a reasonable value of the reduced $-\chi^2$ in their DEM analysis. The analysis was redone without the AIA 131-Å data, and more reasonable results were obtained.

The AIA 131- and 94-Å passbands are particularly problematic since no solar spectrometer has observed these wavelength ranges in decades (*e.g.* Kastner, Neupert, and Swartz, 1974; Acton *et al.,* 1985). Therefore, Testa, Landi, and Drake (2012) used high-resolution X-ray spectra of the low-activity solar-like corona of Procyon obtained with the *Low Energy Transmission Grating Spectrometer* onboard *Chandra*. They found that there may be a significant number of missing low-temperature lines in both AIA channels, so that the AIA flux for a Procyon-like source would be underestimated by the current CHIANTI spectral model by a factor of roughly three in the 94-Å passband and 1.9 in the 131-Å passband. The discrepancy is reduced to about 1.8 for the 94-Å passband when the two Fe IX lines at 93.59 Å and 94.07 Å identified by Lepson *et al.* (2002) are included in the calculation of the AIA response function (Foster and Testa 2011). These results are significantly smaller than the empirical correction factor of 6.7 ± 1.7 for the cool end of the 94-Å response estimated by Aschwanden and Boerner (2011), who used an isothermal approximation.

Del Zanna, O'Dwyer, and Mason (2011) analyze simultaneous observations of ≈ 1 MK loops obtained with AIA and the *Hinode/EUV Imaging Spectrometer* (EIS: Culhane *et al.,* 2007). They use both a loop footpoint and a loop leg region to describe the spectral lines which contribute to the six AIA channels, both directly (171, 193, and 211 Å) and predicted with DEM modeling. They found good overall agreement between observed and predicted count rates for the 131-, 193-, and 335-Å bands. They report that a large number of lines remain unidentified in certain AIA channels. They also extended the work of O'Dwyer *et al.* (2010) related to the multi-thermal nature of the AIA bands to show that emission from lines formed at typical transition region temperatures ($\log T \approx 5.0 - 5.8$) can be significant and even dominant in some cases. They caution that the AIA DEM analysis is unreliable, in agreement with the results of Schmelz *et al.* (2011a,b), and stress the need for further work on the relevant atomic physics before the AIA data can be used routinely for plasma diagnostics.

In this article, we expand on the work described above using simultaneous AIA and EIS observations of X-ray bright points (BPs). BPs were first observed by Vaiana *et al.* (1973) and later revealed to be collections of small coronal loops (Sheeley and Golub, 1979). They are short lived features with sizes ranging from < 15 to ≈ 30 arcseconds and lifetimes of hours to days. BPs are ideal for this work because they are readily observable in both AIA and EIS; they also have narrow temperature distributions that peak near 1 MK, the approximate temperature of the missing lines.

**Observations**

AIA has a full-disk field-of-view, a 1.5-arcsecond spatial resolution, and a 12-second temporal resolution. It consists of four normal incidence multilayer telescopes that provide narrow band imaging in six coronal bandpasses centered on specific iron lines: Fe VIII (131 Å), Fe IX (171 Å), Fe XII (193 Å), Fe XIV (211 Å), Fe XVI (335 Å), and Fe XVIII (94 Å). The temperature diagnostics of the EUV emissions cover the range from 0.4 to 20 million K. Two of the filters also contain hot lines normally associated with flares: the 131-Å channel includes Fe XX and Fe XXIII lines with peak formation temperatures of $\log T = 7.0$ and 7.2, respectively; the 193-Å channel includes Fe XXIV with $\log T = 7.3$. These lines are part of the AIA response functions, but will not affect our results since the emission measure of any hot plasma ($\log T \geq 7.0$) from BPs is likely to be insignificant except in the case of flares (Golub *et al.,* 1974; Strong *et al.,* 1992). Boerner *et al.* (2012) presented the initial photometric calibration of AIA, which was based on preflight measurements of the telescope components. They also described the characterization of the instrument performance and calculated the response functions.

EIS observes the Sun's corona and upper transition region at wavelength ranges of 170 – 210 Å and 250 – 290 Å. Photons enter the EIS instrument through an aluminum filter, which limits the amount of light from optical wavelengths, and are reflected off the primary mirror. The mirror focuses the incoming radiation onto one of four interchangeable slits, with 1, 2, 40, and 266 arc sec widths. The photons are then diffracted onto two back-illuminated CCDs, one for each wavelength range. Images are constructed by rastering across the region of interest and taking exposures at each position. The instrument has a spectral resolution of 47 mÅ at 185 Å and a spatial resolution FWHM of three – four arcseconds.

For the analysis presented here, we selected six BPs observed on three different days from June to September 2010. Figure 1 shows the AIA 193- and 211-Å images where our BPs are marked. The areas selected for analysis are either four × four or five × five EIS pixels (not the entire region inside each box). Since BPs can flicker and even flare, care must be taken to obtain AIA data that are simultaneous with the EIS observations. This is possible since AIA takes data quickly and continuously. We simply need to determine the exact time when the EIS slit is crossing the BP. The dates and times of the EIS and AIA data sets are listed in Table 1.

Standard SolarSoft programs were used to calibrate the data. The EIS data sets contain multiple iron lines of different ionization stages from Fe VIII to Fe XVI. Table 2 lists these lines, their wavelengths in Angstroms, and the logarithm of their peak formation temperatures. EIS lines were fit using eis_auto_fit with intensities in ergs cm$^{-2}$ s$^{-1}$ ster$^{-1}$. The AIA data from the six coronal filters were normalized by the appropriate exposure time, resulting in units of data numbers (DN) sec$^{-1}$. Atomic physics data were obtained from the CHIANTI version 7 (Dere *et al.*, 1997; Landi *et al.*, 2012) data base. The emissivity functions for the EIS lines and the response functions for the AIA filters used the coronal abundances of Schmelz *et al.* (2012b), the update of the hybrid abundances of Fludra and Schmelz (1999), and the ionization balance calculations of both Mazzotta *et al.* (1998) and Bryans, Landi, and Savin (2009).

The data analyzed here were selected from EIS images of quiescent BPs with no significant emission in the Fe XV line at 284 Å or the Fe XVI line at 263 Å, with peak formation temperatures of log $T$ = 6.3 and 6.4, respectively. This selection criterion allows us to use the Fe VIII line at 185.21 Å (where available), which has a significant Ni XVI blend at hotter temperatures. It is important to note that very cool transition region lines that might contribute to the AIA responses are not part of this analysis; our coolest line is Fe VIII. For more information on possible transition region contributions to the AIA filters, please see, *e.g.*, Del Zanna, O'Dwyer, and Mason (2011).

**Analysis**

We use two different DEM methods in our multi-thermal analysis. The first, DEM_manual (Schmelz *et al.*, 2010, 2011a), is a forward folding technique with a manual manipulation of the DEM. The best fit is determined from a $\chi^2$ minimization of the differences between the observed and predicted line fluxes. The main advantages of this method are that no smoothing is required beyond that imposed by the temperature resolution of the spectral line emissivity functions or AIA response functions. Also, no *a-priori* shape (Gaussian or double Gaussian, for example) is imposed on the final DEM curve. The main disadvantage of DEM_manual is that it is not usually possible to explore a broad parameter space, so families of solutions might be missed.

Our second DEM method, xrt_dem_iterative2, is a program first developed (Weber *et al.*, 2004) and tested (Schmelz *et al.*, 2009) for XRT data alone, but now applied more generally (*e.g.* Schmelz *et al.*, 2010; Winebarger *et al.*, 2011). The routine employs a forward-fitting approach where a DEM is guessed and folded through each response to generate predicted fluxes. This process is iterated to minimize $\chi^2$ for the predicted-to-observed flux ratios. The DEM function is interpolated using $N_i$ - 1 splines, representing the degrees of freedom for $N_i$ observations. This routine uses Monte-Carlo iterations to estimate errors on the DEM solution. For each iteration, the observed flux in each line/filter was varied randomly and the program was run again with the new values. The distribution of these variations was Gaussian with a centroid equal to the observed flux and a width equal to the uncertainty.

The intensity for each spectral line observed by EIS is proportional to $\Sigma\ G(T) \times \mathrm{DEM}(T)\ \Delta T$, where $G(T)$ is the emissivity function (erg cm$^3$ s$^{-1}$) from CHIANTI and DEM is the differential emission measure (cm$^{-5}$ K$^{-1}$). The intensity for each AIA filter is proportional to $\Sigma\ \mathrm{Resp}(T) \times \mathrm{DEM}(T)\ \Delta T$, where Resp($T$) is the response function (DN s$^{-1}$ pixel$^{-1}$ per unit emission measure) calculated from the instrument effective areas available in SolarSoft and the CHIANTI synthetic solar spectrum with the same abundance and ionization equilibrium assumptions used for the spectral lines. The AIA and EIS were co-aligned, and the different temporal and spatial resolutions were taken into account. Note that the EIS and AIA units do not have to be converted to the same units, as long as the proper emissivity and response functions are used.

We used the EIS Fe VIII–Fe XVI lines and the plasma density determined from the Fe XII 186.88-to-195.12-Å pair to produce a DEM curve for each BP. We then used these curves and the corresponding AIA 193-Å observed intensity, the channel considered the most reliable, to determine an average EIS–AIA cross-calibration factor for the six BPs. One important caveat, however, needs to be addressed before we proceed to the final BP DEM analysis. Since Fe VIII is the coolest line/filter in the BP data sets, the DEM curve can always be adjusted to produce a mathematically acceptable Fe VIII fit by simply adding enough low-temperature plasma (without affecting the Fe IX line/filter). An example from the AIA loop analysis of Schmelz *et al.* (2011b) is shown in Figure 2 for their loop (g). The top left panel shows the DEM with the low-temperature emission required to produce a good mathematical fit to the data. The top right panel shows the predicted-to-observed intensity ratios for the six AIA coronal filters with reduced-$\chi^2$ = 1.0. The problem is that the loop data have been background subtracted, so the cool transition region plasma should not be part of a physically acceptable model. A simple DEM model and the resulting predicted-to-observed intensity ratios are shown in the bottom panels. The ratio for the 131-Å (Fe VIII) channel is too low, and drives up the reduced-$\chi^2$ = 18.5, far too high for a mathematically acceptable solution. The problem with the AIA 131-Å (Fe VIII) channel must have another solution; searching for that solution is the main purpose of this article.

With this important caveat in mind, we have elected to consider only simple DEM curves for the BP analysis. These results are shown in Figure 3, which use both the EIS and AIA data. These results are from xrt_dem_iterative2, where the solid histogram is the minimum-$\chi^2$ result, and the dashed red histograms are the Monte Carlos.

Figure 4 shows the results using the coronal abundances of Schmelz *et al.* (2012) and the ion fractions of Mazzotta *et al.* (1998) to calculate both the EIS emissivity functions and the AIA response functions. Note that since we use only EIS iron lines and the AIA coronal filters are dominated by iron lines, the abundances should not affect our results as long as the same set is used throughout, which is indeed the case. Each panel shows the predicted-to-observed intensity ratios for the EIS lines and the AIA channels, with the data points plotted in order of their peak formation temperature (see Table 2). The symbols shown in black represent no alterations to the AIA response functions. The red data points show the results if the AIA 131-Å cool-temperature response curve is multiplied by a factor of 1.9, a value consistent with the Testa, Landi, and Drake (2012) results for Procyon. The blue data points show the same thing, except the AIA 131-Å cool-temperature curve is multiplied by a factor of $2 \times 1.9$. Note that although the AIA 94-Å channel is labeled as Fe XVIII, the bright point emission is most likely dominated by the cooler temperature Fe IX and Fe X lines.

The largest discrepancies are for Fe VIII and Fe IX (on the left side of each panel). It is possible to obtain a DEM curve that does a reasonable-to-good job of reproducing the observed fluxes for one or the other, but not for both. BP (a) from 8 June 2010 has no EIS Fe VIII or Fe IX lines to help us understand this discrepancy. BPs (b), (c), and (d) from 25 June 2010 have EIS Fe VIII 186.6 Å and Fe IX 188.5 and 197.9 Å. BPs (e) and (f) from 3 September 2010 have Fe VIII 185.2 and 186.6 Å and Fe IX 197.9 Å. These EIS lines can help us pin down the scope of the problem.

In each case, the predicted-to-observed intensity ratios for the Fe VIII EIS lines and Fe VIII AIA 131-Å filter are significantly below one. Increasing the emission measure at $\log T \approx 5.6$, the peak formation temperature, will move these point up, but it will also raise the Fe IX EIS lines and Fe IX AIA 171-Å filter, which are already high enough. We know that there is a significant number of missing low-temperature lines in Fe VIII AIA 131-Å channel. If we attempt to account for these missing lines using the results of Testa, Landi, and Drake (2012), we can reduce the discrepancy for AIA (red and blue data points), but this does nothing to change the position of the EIS lines.

We repeated the DEM_manual and xrt_dem_iterative2 analysis with the Bryans, Landi, and Savin (2009) ion fractions. These values are used to construct the default AIA response functions available in SolarSoft. The changes from Mazzotta *et al.* (1998) to Bryans, Landi, and Savin (2009) were particularly dramatic for Fe VIII and Fe IX, the ions that were most discrepant in the ratio plots of Figure 4. In fact, the peak formation/response temperature for Fe VIII moves from $\log T = 5.6$ to 5.7. Our new results are shown in Figure 5, which has the same format as Figure 4. Although the discrepancies between Fe VIII and Fe IX diminish and the overall results improve, we were still not able to get acceptable fits for the DEM analysis for any of our BPs. This was true even when we scaled up the AIA response functions to account for cool lines known to be missing from the CHIANTI data base (red and blue data points). Note that the predicted-to-observed intensity ratios for the Fe VIII EIS lines are still too low in each panel.

The results seen in Figures 4 and 5 suggested that the problem encountered here might be related to the ionization balance. It might be possible that the changes from Mazzotta *et al.* (1998) to Bryans, Landi, and Savin (2009) for Fe VIII and Fe IX did not go far enough. In other words, could later updates to the Fe VIII ionization equilibrium move the peak formation/response temperature to slightly higher values, from $\log T = 5.7$ to 5.75 or 5.8?

Many of the ionization rates have been measured in the laboratory. Bryans, Landi, and Savin (2009) used the Dere (2007) cross sections, which are similar to those used by Arnaud and Raymond (1985) and Mazzotta *et al.* (1998). The Fe VII and Fe X ionization rates were measured with estimated uncertainties of 8 %. This is not true for Fe VIII or Fe IX, however. For these, Dere (2007) interpolated using Mn and Ni measurements. Given the often poor agreement between measured and calculated cross sections, the uncertainty could be at least 20 % (J. Raymond, private communication, 2011).

The recombination rates have changed even more, mainly because these are complex ions and dielectronic recombination is sensitive to the model of the ion. Nikolíc *et al.* (2010) find a somewhat higher rate for Fe IX dielectronic recombination than earlier calculations, which is 40 % in the relevant temperature range and even higher at very low temperatures. The measured values are even higher still. There are as of yet no compilations that use the new dielectronic recombination rates, but the magnitude of the expected changes should be similar (J. Raymond, private communication, 2011).

As a result of these uncertainties, we manually shifted the cool peak of the AIA 131-Å response as well as the EIS Fe VIII emissivity functions from $\log T = 5.7$ to 5.75 and then to 5.8. We find that $\log T = 5.75$ is not quite enough, but the DEM results are quite good for $\log T = 5.8$. We once again adjusted this peak to bring our BP DEM results into better agreement. The weighted mean of the data from the six BPs indicates that we need to multiply the AIA 131-Å cool-temperature peak by a factor of 1.25 for the best DEM results. This factor is slightly smaller than the one found by Testa, Landi, and Drake (2012) for the Procyon data, but they used the ionization equilibrium calculations of Bryans, Landi, and Savin (2009). The final results are illustrated in Figure 6, and show that the Fe VIII and Fe IX ratios can now be reconciled for all six BPs. (Note that there would also be a corresponding 10 – 20 % reduction to the Fe IX ionization equilibrium calculations at $\log T \approx 5.8$. A more precise value will result from updated laboratory measurements and calculations of the cross sections and recombination rates.)

The cool-temperature peak of the AIA 94-Å response required no changes in ionization equilibrium, but we did adjust this peak to bring our BP DEM results into better agreement. The weighted average of the data from the six BPs indicates that we need to multiply the AIA 94-Å cool-temperature peak by a factor of 1.72, which is in good agreement with the results of Foster and Testa (2011) and slightly smaller than the scaling factor found by Testa, Landi, and Drake (2012) for the Procyon data. These results are illustrated in Figure 6.

Our ability to pinpoint the problem to the Fe VIII ionization equilibrium calculations comes in part from the substantial improvements made recently in the atomic data for the iron lines used in this analysis (*e.g.* Del Zanna *et al.*, 2004; Storey *et al.*, 2005; Young, 2009). If the predicted-to-observed intensity ratios for the EIS Fe IX–Fe XIII lines were not so close to one, as seen in Figures 3 – 5, then the problem affecting the AIA 131-Å data would have been much more difficult to track down.

**Discussion**

Figures 4 and 5 show that accounting for the missing lines observed in the Procyon spectrum (Testa, Landi, and Drake, 2012) can improve the results for the AIA 131-Å CHIANTI data, but these changes alone are not enough to explain the entire problem. The discrepancies between the Fe VIII and Fe IX BP data cannot be resolved simply by accounting for the missing lines, especially when the AIA and EIS data are analyzed together. Our results show that the predicted-to-observed intensity ratios for the EIS 185.21- and 186.60-Å Fe VIII lines are always ≤ 0.2 for Mazzotta *et al.* (1985) and ≤ 0.4 for Bryans, Landi, and Savin (2009). A more complete CHIANTI data base is unlikely to resolve these discrepancies. Our results suggest that the problem encountered here might be related to the ionization balance. Empirical tests with the BP EIS–AIA data suggest that the changes from Mazzotta *et al.* (1998) to Bryans, Landi, and Savin (2009) for Fe VIII did not go far enough and that the peak temperature of the Fe VIII emissivity/response should move to a slightly higher value, to $\log T = 5.8$. Our suggestion for a revised AIA 131-Å response function appears in Figure 7a, where the cool-temperature peak shown in red was multiplied by a factor of 1.25 to account for missing lines. To summarize, a combination of missing spectral lines and a small change to the Fe VIII ionization fraction can resolve the discrepancy between the AIA–EIS observations and the predicted DEM intensities.

The issue of the Fe VIII ion fractions was discussed by Brooks, Warren, and Young (2011), who found that the atomic calculations from Dere *et al.* (2009) were in agreement with EIS observations. Their work used CHIANTI 6, however, and the Fe VIII model was updated in

CHIANTI 7. Based on the assessment of Fe VIII by Del Zanna (2009), the collision strengths for both the 185.21- and 186.60-Å Fe VIII lines were scaled by about 50 %, which uses the more accurate A-value calculations to help improve the collision data. For more details, please see Young and Landi (2009). As a result, more emission measure would be required to obtain acceptable predicted-to-observed intensity ratios for these lines, which is consistent with our results.

We find that the AIA 171-Å channel data are in reasonable agreement with the EIS Fe IX 188.5- and 197.9-Å lines. With the exception of BP (f), the results are within the measured uncertainties. To the extent that there is disagreement, the predicted-to-observed intensity ratio for the AIA 171-Å data is high, the opposite of what would be expected if the CHIANTI data base were incomplete in this wavelength range. It is worth pointing out that our result is not in agreement with that of Del Zanna, O'Dwyer, and Mason (2011). Their analysis of AIA–EIS data results in a loop footpoint DEM (their Figure 6) that peaks near $\log T = 5.7$, the peak formation temperature of Fe VIII and, therefore, the peak of the cool end of the AIA 131-Å response. In addition, their DEM dips at $\log T = 5.9$, the peak of the AIA 171-Å response. We attempted to fit our BP data with a similar curve, like the one seen in Figure 2, but this did not solve the Fe VIII–Fe IX problem. Our analysis uses only iron lines, which is both an advantage because we do not have to assume a set of elemental abundances, and a disadvantage because Fe VIII is our lowest temperature line. Del Zanna, O'Dwyer, and Mason (2011), on the other hand, use lines from iron, silicon, magnesium, and oxygen. A direct comparison with EIS is not necessarily straightforward because the AIA 171-Å band is at the edge of the EIS sensitivity. In at least one of their examples (the loop footpoint), however, the AIA observations and the EIS predictions are in excellent agreement, but the DEM predictions are too high. All of these results are consistent with the conclusion that the CHIANTI data base is reasonably complete near 171 Å, at least for coronal temperatures. To summarize, part of the AIA 131- and 171-Å discrepancy is certainly related to the Fe IX ionization balance, but we recommend that the 171-Å response function remain unchanged until updated Fe IX ionization equilibrium measurements and calculations are complete.

Since our BP DEM results were cross-calibrated using the AIA 193-Å channel and the EIS 195-Å Fe XII line, it should come as no surprise that there is excellent agreement. This is also true for the Del Zanna, O'Dwyer, and Mason (2011) results for both the loop leg and the footpoint. The AIA 193-Å channel covers a small wavelength range in the heart of one of the two EIS channels (170 – 210 Å), so direct comparisons are possible. Similar channels on the EIT and TRACE instruments have ensured that this waveband is well-studied. We conclude that the CHIANTI data base is reasonably complete near 193 Å and recommend that the 193-Å response function remain unchanged.

We find that the AIA 211-Å channel data are in reasonable agreement with our BP DEM results. Although all of our BPs are detected in this channel (see Figure 1), the EIS 274.21-Å Fe XIV line was not available in any of these data sets. The EIS 264.79-Å line was available for BP (a) and the 270.52-Å line was available for BPs (e) and (f), but the emission measure at $\log T = 6.3$, the peak formation temperature of Fe XIV, was too low and no reliable spectral line fits resulted from eis_auto_fit. Like the AIA 171-Å channel, the 211-Å channel is at the edge one of the two EIS channels (170 – 210 Å). Therefore, direct comparisons are possible but the sensitivity of EIS is uncertain. The results from Del Zanna, O'Dwyer, and Mason (2011) indicate that a fair number of lines remain unidentified in this band. We recommend that more work be done with AIA–EIS DEM comparisons using observations of active region cores, regions with more warm emission measure than our BPs. Then direct comparisons can be made not only with the EIS 211-Å band, but also with the EIS 264.79-, 270.52-, and 274.21-Å Fe XIV lines.

Our BPs show very little emission in the AIA 335-Å images and we can, therefore, make no independent recommendations for this channel. There is a second order peak in the mirror reflectivity and some cross-talk (Boerner *et al.*, 2012; Del Zanna, O'Dwyer, and Mason, 2011). Results from Del Zanna, O'Dwyer, and Mason (2011) indicate that the CHIANTI data base is reasonably complete near 335 Å. We recommend that more work be done with AIA–EIS DEM comparisons using observations of active region cores. Then direct comparisons can be made with the EIS 262.98-Å Fe XVI line.

Figures 4 and 5 show that accounting for the two Fe IX lines at 93.59 and 94.07 Å identified in the EBIT data by Lepson *et al.* (2002) and observed in the Procyon spectrum by Testa, Landi, and Drake (2012) can improve the BP DEM results for the AIA 94-Å data. Although our BPs are not ideal for pinning down the problems in the 94-Å CHIANTI data — the emission measures are not high enough to give a strong Fe XVIII signal — our results are nonetheless consistent, within the uncertainties. The weighted average of the data from the six BPs indicates that we need to multiply the AIA 94-Å Fe IX/Fe X cool-temperature peak by a factor of 1.72. Our results are shown in Figure 7b. We recommend that the AIA response function calculated by Foster and Testa (2011) be adopted for the AIA 94-Å channel and that more work be done with AIA–EIS DEM comparisons using observations of active region cores.

**Conclusions**

Our DEM analysis of six BPs observed simultaneously by AIA and EIS support the conclusion that the CHIANTI data base used to model the response functions is incomplete near 131 Å. We find that the observed flux is underestimated by $\approx 1.25$. A more important result, however, is that the peak temperature of the Fe VIII emissivity/response is likely to be closer to $\log T = 5.8$ than to the currently accepted value of $\log T = 5.7$ (Bryans, Landi, and Savin, 2009). Making these adjustments brings not only the AIA 131-Å data, but also the EIS 185.21- and 186.60-Å Fe VIII lines, into agreement with the remainder of the AIA–EIS BP data.

In addition, we find that the CHIANTI data base is reasonably complete in the AIA 171- and 193-Å bands and recommend no changes to these response functions. For the AIA 211- and 335-Å channels, we recommend that more work be done with AIA–EIS DEM comparisons using observations of active region cores, regions with more warm emission measure than our BPs. Then direct comparisons can be made with the EIS Fe XIV and Fe XVI lines. Finally, we recommend that the AIA 94-Å response function calculated by Foster and Testa (2011) be adopted for any DEM analysis.

**Acknowledgements**

The authors would like to thank John Raymond and Nancy Brickhouse of CfA and Peter Young and David Brooks of NRL for helpful discussions on Fe VIII. AIA data are courtesy of

NASA/SDO and the AIA science team. *Hinode* is a Japanese mission developed and launched by ISAS/JAXA, with NAOJ as domestic partner and NASA and STFC (UK) as international partners. It is operated by these agencies in co-operation with ESA and the NSC (Norway). CHIANTI is a collaborative project involving the NRL (USA), the Universities of Florence (Italy) and Cambridge (UK), and George Mason University (USA). Solar physics research at the University of Memphis is supported by a *Hinode* subcontract from NASA/SAO as well as NSF ATM-0402729.

Table 1

EIS and AIA Bright Point Data Sets

| ID | Position | Date | EIS Time | log $T$ | $n_e$ [$10^8$ cm$^{-3}$] | AIA Time |
|----|----------|------|----------|---------|--------------------------|----------|
| a | S04E08 | 08 Jun. 2010 | 18:15:41 | 6.08 | 7 | 19:48:16 |
| b | S08E06 | 25 Jun. 2010 | 20:10:41 | 6.13 | 6 | 20:28:41 |
| c | S12E11 | 25 Jun. 2010 | 20:10:41 | 6.16 | 8 | 20:48:32 |
| d | S17E12 | 25 Jun. 2010 | 20:10:41 | 6.14 | 9 | 20:28:41 |
| e | S31E51 | 03 Sep. 2010 | 03:10:41 | 6.04 | 6 | 03:49:41 |
| f | S35E63 | 03 Sep. 2010 | 03:10:41 | 6.03 | 6 | 03:26:11 |

Table 2

EIS Lines from Bright Point Data Sets

|    | Ion     | λ [Å]  | log $T$ |
|----|---------|--------|---------|
| 1  | Fe VIII | 185.21 | 5.6     |
| 2  | Fe VIII | 186.60 | 5.6     |
| 3  | Fe IX   | 188.50 | 5.9     |
| 4  | Fe IX   | 197.86 | 5.9     |
| 5  | Fe X    | 184.53 | 6.0     |
| 6  | Fe XI   | 180.40 | 6.1     |
| 7  | Fe XI   | 188.22 | 6.1     |
| 8  | Fe XII  | 186.85 | 6.1     |
| 9  | Fe XII  | 192.39 | 6.1     |
| 10 | Fe XII  | 193.51 | 6.1     |
| 11 | Fe XII  | 195.12 | 6.1     |
| 12 | Fe XII  | 203.74 | 6.1     |
| 13 | Fe XIII | 202.04 | 6.2     |
| 14 | Fe XIII | 203.80 | 6.2     |
| 15 | Fe XIII | 203.83 | 6.2     |
| 16 | Fe XIV  | 264.79 | 6.3     |
| 17 | Fe XIV  | 270.52 | 6.3     |
| 18 | Fe XV   | 284.16 | 6.3     |
| 19 | Fe XVI  | 262.98 | 6.4     |

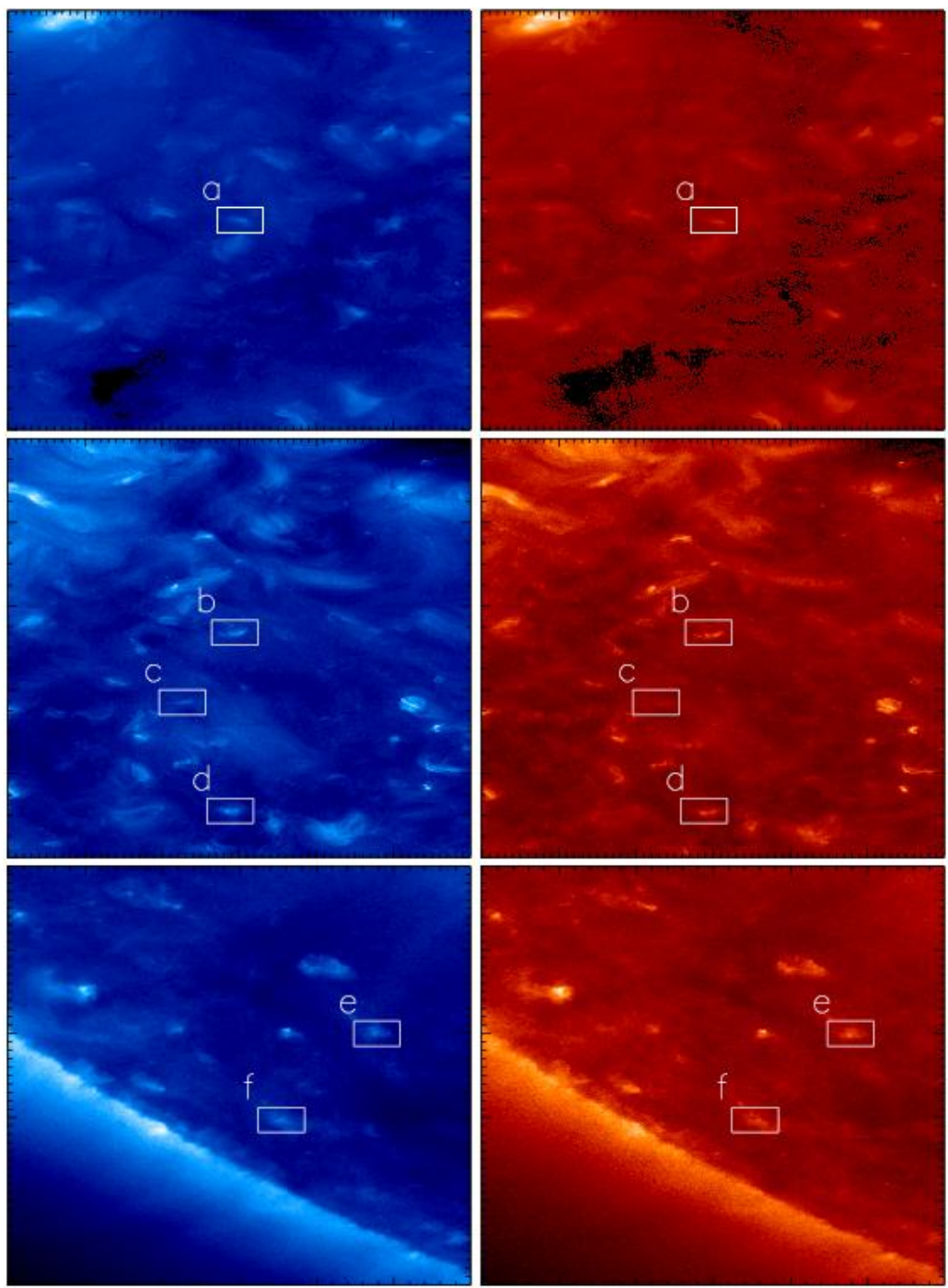

Figure 1
AIA 193- (left) and 211-Å (right) images from 8 June 2010 (top), 25 June 2010 (middle), and 3 September 2010 (bottom). The FOV for each image is 350 × 500 square arc-seconds. The BPs analyzed here are marked by the white boxes. The areas selected for analysis are either 4 × 4 or 5 × 5 EIS pixels at the peak of the emission (not the entire region inside each box).

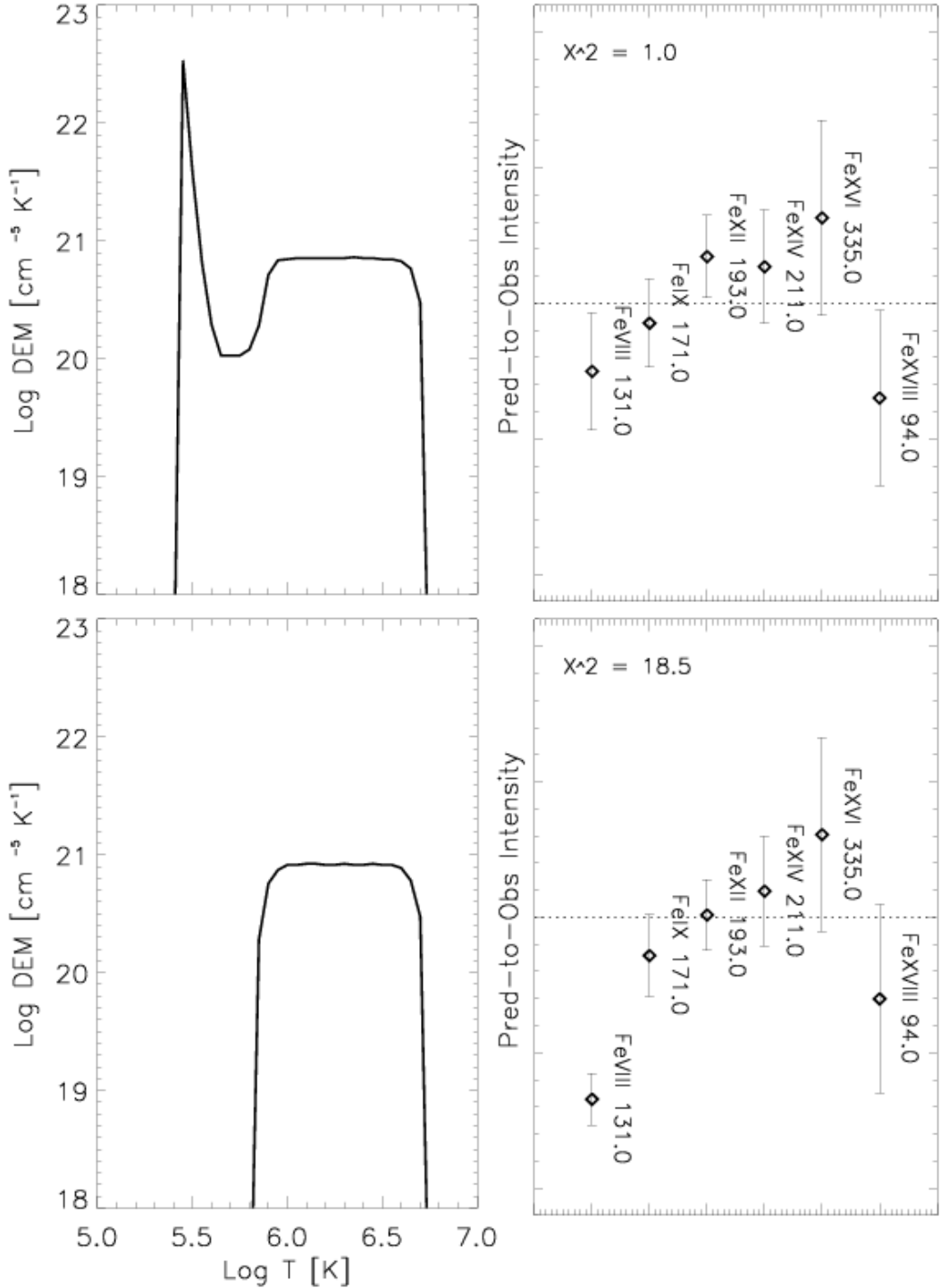

Figure 2
Example from the AIA-loop analysis of Schmelz *et al.* (2011b). The DEM with low-temperature emission required to produce a good mathematical fit to the data (top left); predicted-to-observed intensity ratios (top right) for the six AIA coronal filters with reduced-$\chi^2$ = 1.0.. A simple DEM model (bottom left) and the resulting predicted-to-observed intensity ratios (bottom right) with reduced-$\chi^2$ = 18.5.

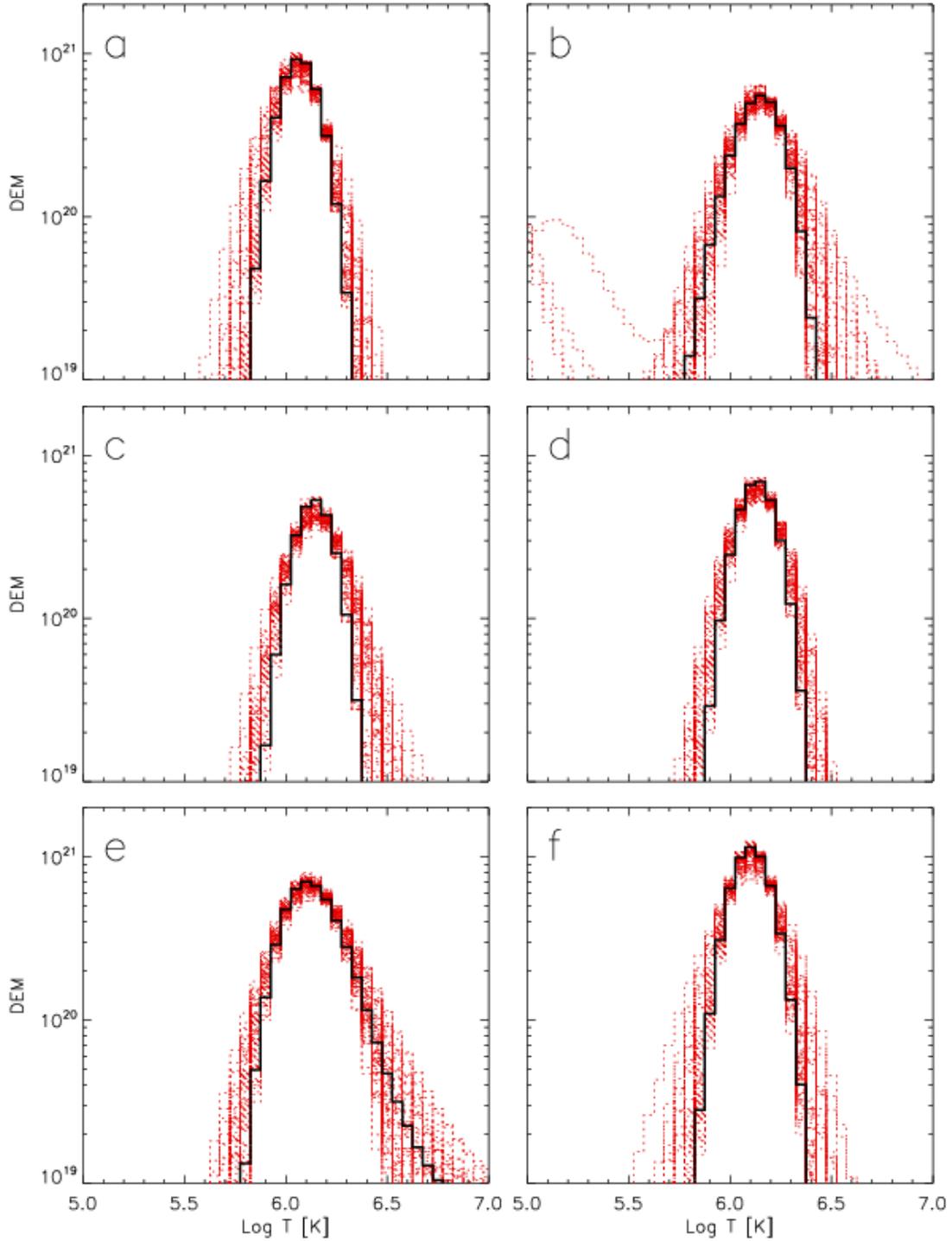

Figure 3
Multi-thermal DEM results from xrt_dem_iterative2 for each BP listed in Table 1 and observed in Figure 1. The solid black histogram is the minimum-$\chi^2$ result. Each dashed red line shows a single Monte-Carlo realization. The tightness of each distribution shows the good DEM fit to the data.

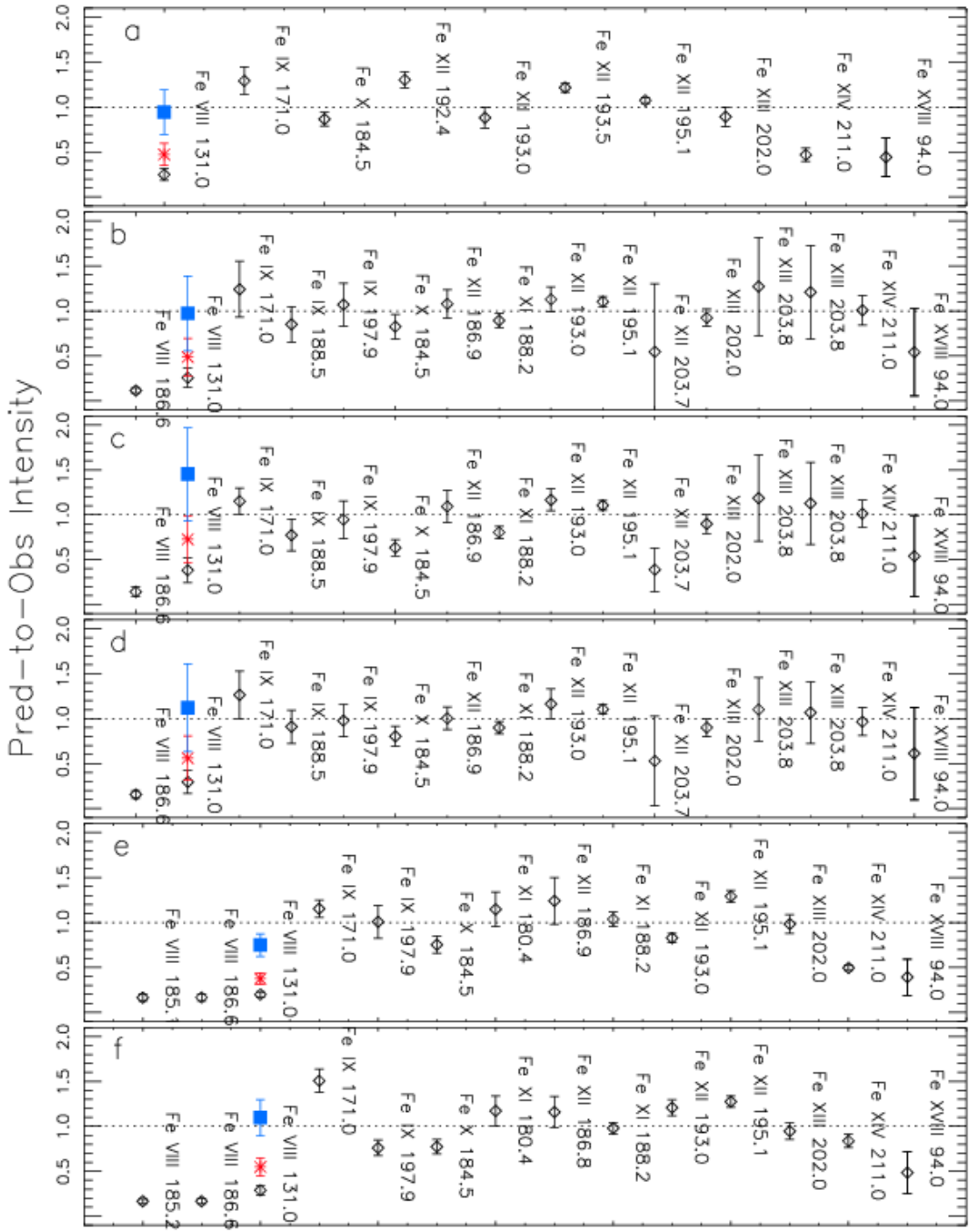

Figure 4
Predicted-to-observed intensity ratio for each EIS line and AIA channel using the Mazzotta *et al.* (1998) ion fractions. Each panel shows the results for a different BP. The black diamonds represent no alteration to the AIA response function. The red star and blue square show the results if the AIA 131-Å cool-temperature response curve is multiplied by a factor of 1.9 and 2 × 1.9, respectively (see text). The reduced-$\chi^2$ values are (a) 27, 16, 14; (b) 66, 63, 63; (c) 20, 19, 19; (d) 26, 25, 25; (e) 136, 100, 93; (f) 148, 137, 135; for the black, black plus red, and black plus blue data points, respectively.

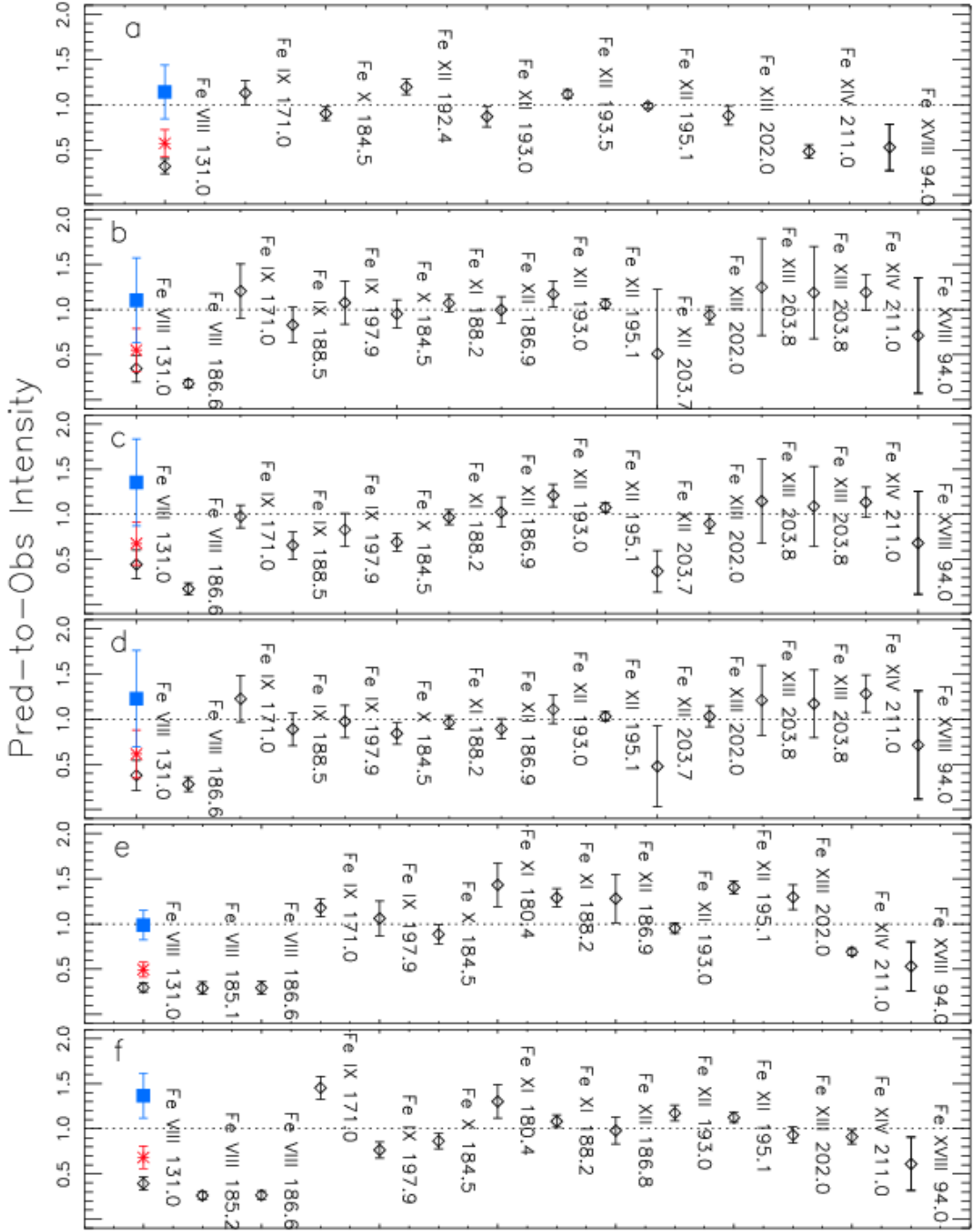

Figure 5
Predicted-to-observed intensity ratio for each EIS line and AIA using the Bryans, Landi, and Savin (2009) ion fractions. Note the differences for Fe VIII and Fe IX, the data points on the left side of each panel. The reduced-$\chi^2$ values are (a) 17, 11, 10; (b) 23, 22, 22; (c) 12, 11, 11; (d) 7.0, 6.2, 6.0; (e) 40, 28, 25; (f) 49, 44, 43; for the black, black plus red, and black plus blue data points, respectively.

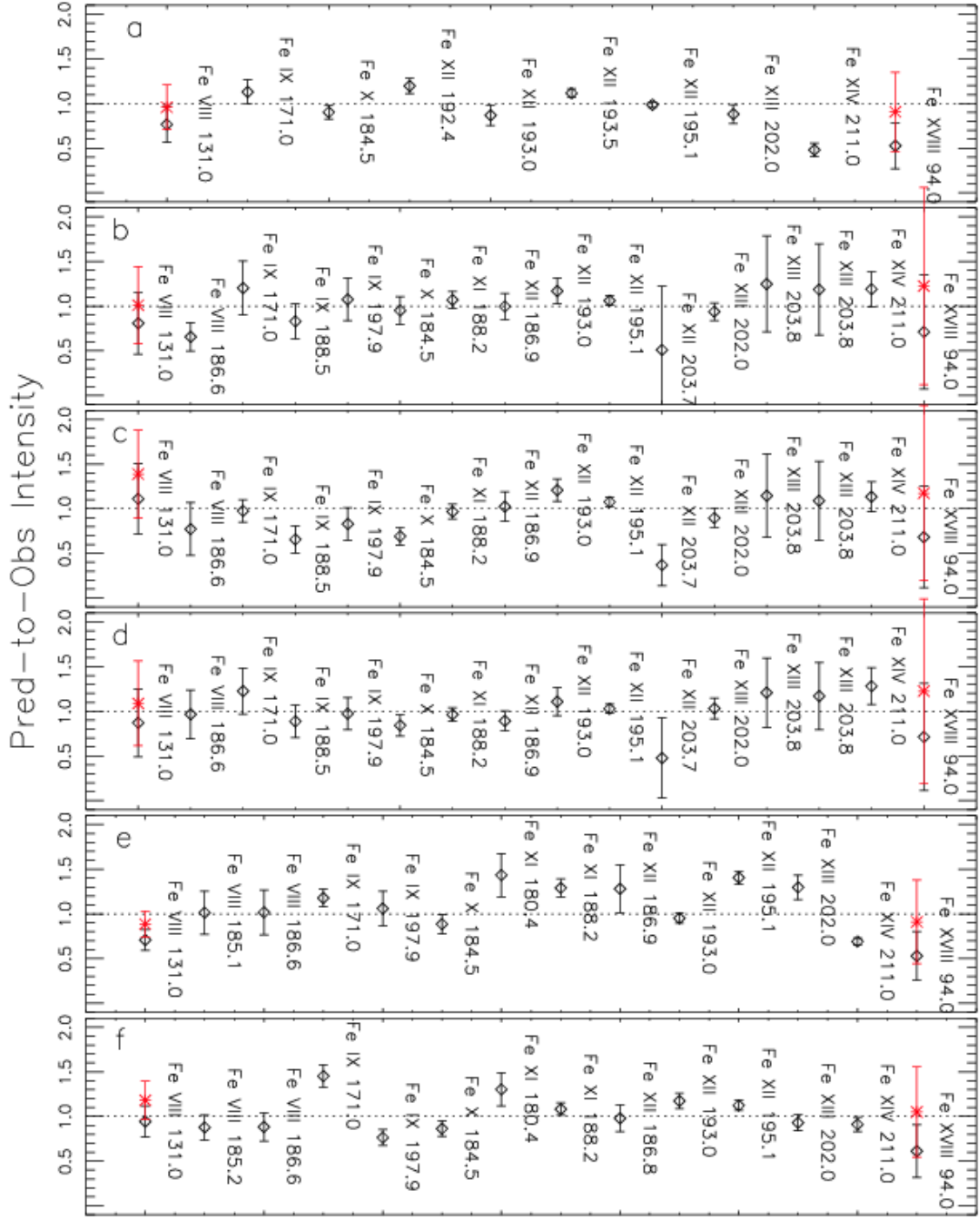

Figure 6
Predicted-to-observed intensity ratio for each EIS line and AIA using the Bryans, Landi, and Savin (2009) ion fractions after shifting the cool peak of the AIA 131-Å response as well as the EIS Fe VIII emissivity functions from log $T = 5.7$ to $5.8$. The red data points show the results if the AIA 131- and 94-Å cool-temperature response curves are multiplied by factors of 1.25 and 1.72, respectively, the weighted mean of the data from the six BPs (see text). The reduced-$\chi^2$ values are (a) 7.4, 6.9; (b) 1.0, 1.0; (c) 2.0, 2.0; (d) 1.0, 1.0; (e) 9.8, 9.2; (f) 3.1, 3.0; for the black and black plus red data points, respectively.

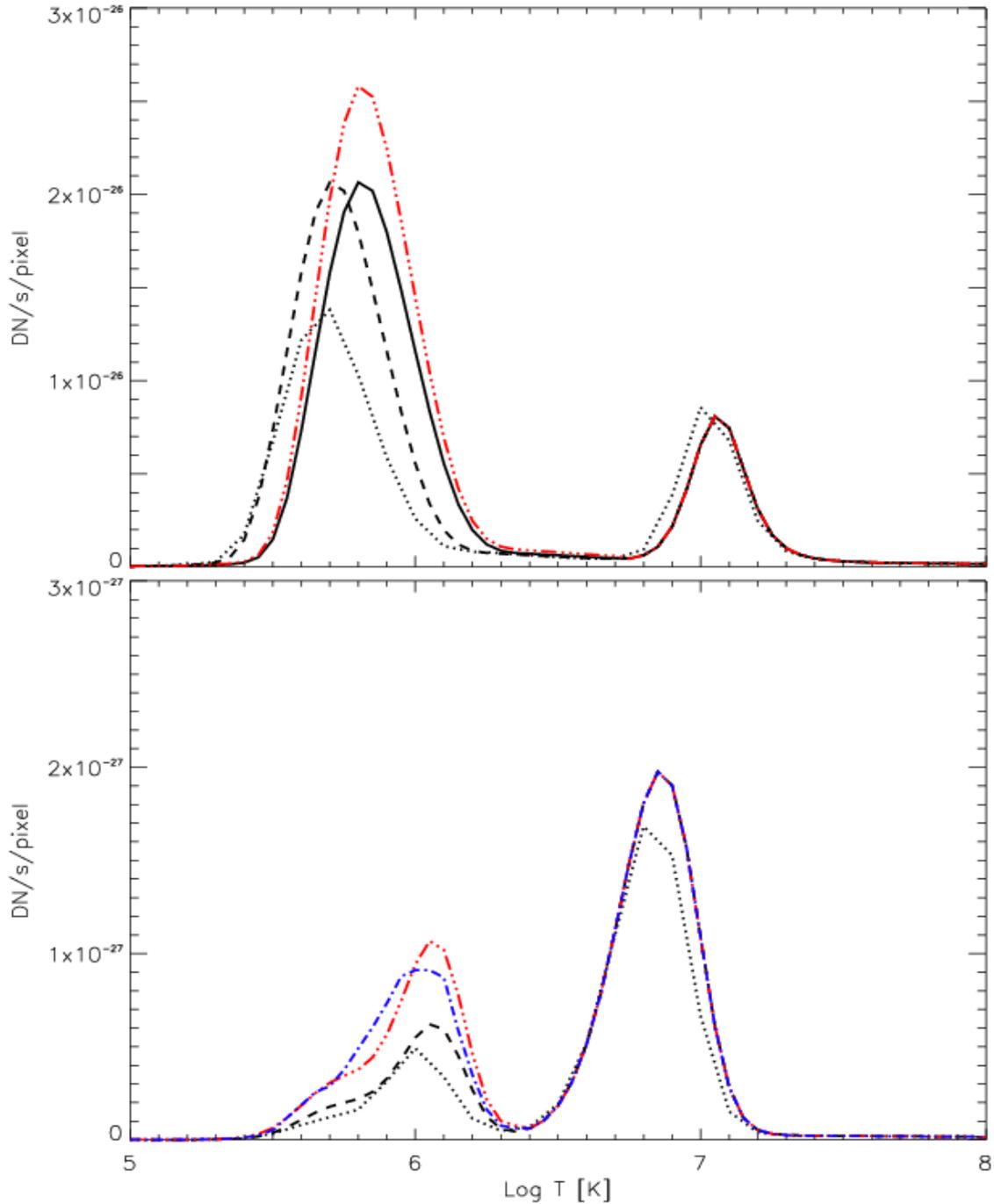

Figure 7
(a) AIA 131-Å response functions assuming ionization fractions from Mazzotta *et al.* (1997) (dotted), Bryans *et al.* (2009 (dashed), Bryans *et al.* (2011) but with the cool-temperature peak adjusted manually from log $T$ = 5.7 to 5.8 (solid), and then enhanced (red) by a factor of 1.25 to account for missing lines (see text). (b) AIA 94-Å response functions assuming ionization fractions from Mazzotta *et al.* (1997) (dotted), Bryans *et al.* (2009) (dashed), Bryans *et al.* (2009) enhanced (red) by a factor of 1.72 to account for missing lines (see text). For comparison, the modified response from Foster and Testa (2011) is shown in the dot–dashed blue curve.